\documentclass[preprintnumbers, amsmath, amssymb, showpacs,twocolumn, superscriptaddress, 
nofootinbib]{revtex4}

\pdfoutput=1

\usepackage{graphicx,subfigure,bm,color,psfrag,hyperref}
\usepackage{amsfonts}
\usepackage{lipsum}
\usepackage{mathtools}
\usepackage{verbatim}

\newcommand{\be}{\begin{equation}}
\newcommand{\ee}{\end{equation}}

\begin{document}

\title{New observational constraints on $f(T)$ gravity through gravitational-wave 
astronomy}

\author{Rafael C. Nunes}
\email{rafadcnunes@gmail.com}
\affiliation{Divis\~ao de Astrof\'isica, Instituto Nacional de Pesquisas Espaciais, 
Avenida dos 
Astronautas 1758, S\~ao Jos\'e dos Campos, 12227-010, SP, Brazil}

\author{Supriya Pan}
\email{supriya.maths@presiuniv.ac.in}
\affiliation{Department of Mathematics, Presidency University, 86/1 College Street, 
Kolkata 700073, India}

\author{Emmanuel N. Saridakis}
\email{Emmanuel\_Saridakis@baylor.edu}
\affiliation{Department of Physics, National Technical University of Athens, Zografou
Campus GR 157 73, Athens, Greece}
\affiliation{CASPER, Physics Department, Baylor University, Waco, TX 76798-7310, USA}

\begin{abstract}
We investigate the new observational constraints on $f(T)$ gravity that arise from  the 
effects of primordial gravitational waves (GWs) on the cosmic microwave background (CMB) 
anisotropies and the BB spectrum. We first show that  on the GWs propagation in $f(T)$ 
gravity we obtain 
only an amplitude modification and not a phase one, comparing to 
the case of general relativity in the background of $\Lambda$CDM cosmology. Concerning 
primordial GWs  we find that  the more the model departs from general relativity  the 
larger is the GW amplitude decay, and thus a possible future detection  would  bring the 
viable $f(T)$ gravity models five orders of magnitude closer to $\Lambda$CDM cosmology 
comparing to standard cosmological constraints. Additionally, we use the CLASS code and 
both   data from the Planck probe, as well as forecasts from the near-future CORE 
collaboration, and we show that  possible non-trivial constraints on the tensor-to-scalar 
ratio would offer a clear signature of $f(T)$ gravity. Finally, 
we discuss on the possibility to use the properties of the GWs that arise from neutron 
stars  mergers in order to extract additional constrains on the theory.

\end{abstract}


\maketitle
\section{Introduction}

The LIGO and Virgo collaborations reported the first direct detection of gravitational 
waves (GWs) through the GW150914 event \cite{ligo01}, that was produced from the merger 
of 
a pair of black holes with masses around 36 and 29 solar masses and the subsequent 
ringdown of a single black hole, while similar GWs event were later reported 
in  \cite{Abbott:2016nmj,Gw03,Gw04,Gw05,Gw06}. Recently, the detection of a binary 
neutron star merger with a GW (GW170817 event \cite{Gw07}) and an electromagnetic 
counterpart (GRB 170817A event \cite{GRB17}) opened the window of multi-messenger GW 
astronomy.
Certainly, the detection of GWs is a potential indication for a new era in modern 
astrophysics and cosmology, offering a new spectrum of possibilities to 
investigate nature at the fundamental level.

On the other hand, another important aspect in cosmology is to  investigate the cosmic 
origin through a possible detection of primordial gravitational waves and their effect 
on the Cosmic Microwave Background (CMB) in the very low frequency band, which is 
expected to bring new astrophysical and cosmological information. 
The primordial GWs can be quantified through the tensor-to-scalar ratio $r$ evaluated on 
some pivot scale. The Planck collaboration  within $\Lambda$CDM$+r$ 
has reported $r < 0.10$ at 95\% confidence level (CL) by combining temperature, 
low-polarization, and lensing, at the pivot scale $k_{*} = 0.002 \, \, {\rm Mpc^{-1}}$  
\cite{Akrami:2018odb}.

Concerning the research on the fundamental gravitational interaction, the GWs detection 
can be extremely helpful since the underlying gravitational theory determines both the 
properties of the GWs themselves (speed, polarization modes etc) as well as the 
properties of the background on which they propagate (the expanding universe). Thus, one 
can use the GWs measured properties and extract valuable information for the structure of 
the fundamental gravitational interaction. For instance, from the recent GW170817 
and GRB 170817A events, which showed that GWs propagate practically with the light 
speed, one may impose strong constraints or exclude various modified gravity theories 
of the literature \cite{GW_MG01,GW_MG02,GW_MG03,Ezquiaga:2017ekz} (see \cite{GW_MG05} 
for a latest review). Implications of an electromagnetic counterpart measurement  for
a range of modified gravity theories was also previously discussed in \cite{GW_MG06, GW_MG07}. 

One class of theories of modified gravity that have recently attracted the interest 
of the research is the torsional gravitational modification, such as the $f(T)$ gravity.
In this theory one describes the gravitational interaction through the torsion instead 
of the curvature tensor, and thus the Lagrangian is a function of the torsion scalar $T$ 
(see \cite{Cai:2015emx} for a review). Hence, $f(T)$ gravity corresponds to a novel 
gravitational modification, with no known curvature equivalence, and it proves to lead to 
interesting cosmology at both  early and  late times
\cite{Ferraro:2006jd,Bengochea:2008gz,Linder:2010py,Chen:2010va,Dent:2011zz,
Zheng:2010am,Zhang:2011qp,Cai:2011tc, Capozziello:2011hj,Geng:2011aj,Wu:2011kh, 
deHaro:2012zt,
Guo:2015qbt,Atazadeh:2011aa, Basilakos:2013rua,Ong:2013qja,Amoros:2013nxa,
Bamba:2013jqa,Paliathanasis:2014iva,Kofinas:2014daa,Geng:2014nfa,Hanafy:2014ica,
Capozziello:2015rda, Malekjani:2016mtm,Farrugia:2016qqe,
Sk:2017ucb,Bahamonde:2017wwk,
Hohmann:2018rwf,Ilijic:2018ulf,Golovnev:2018wbh,Keskin:2018gev,Deng:2018ncg,
El-Zant:2018bsc} and be in agreement with observations
\cite{Wu:2010mn,Nesseris:2013jea,Nunes:2016qyp,Nunes:2016plz,Nunes:2018xbm,
Basilakos:2018arq}.

Some preliminary investigations concerning the GWs properties in $f(T)$ gravity have been 
performed in \cite{Cai:2018rzd,Li:2018ixg}, where it was shown that their speed is equal 
to the light speed and thus the constraints from  GW170817  are trivially satisfied (see 
also \cite{Farrugia:2018gyz,Hohmann:2018wxu} for the GWs properties in other modified 
teleparallel gravities). In this work we are interested in studying the new observational 
constraints that can be imposed on $f(T)$ gravity from the advancing multi-messenger GW 
astronomy. Moreover, we examine the effect which primordial GWs have on the cosmic 
microwave background (CMB) anisotropies and we examine the corresponding constraints. 
Finally, we discuss on the possibility that GWs from neutron star mergers could lead to 
additional constraints on the theory.

The manuscript is organized as follows: In  Section \ref{sec-model} we present the 
equations that determine the propagation of GWs in $f(T)$ gravity, and we quantify their 
modification comparing to the case of general relativity in a background of $\Lambda$CDM 
cosmology. In Section \ref{PrimordialGW} we study the observational constrains that 
arise from primordial GWs, and in Section \ref{PlanckCORE} we use the CLASS code and data 
from the Planck probe and forecasts from the near future CORE collaboration in order to 
quantify the effects of primordial GWs on the  CMB anisotropies and the BB spectrum. In 
Section \ref{mergersGWs} we discuss on the possibility to extract   constraints   
  from GWs that arise from neutron star mergers. Finally, in Section \ref{Conclusions} we 
present the conclusions.

\section{Gravitational waves in $f(T)$ gravity}
\label{sec-model}

In this section we   describe the equations of $f(T)$ gravity, as well 
as the formulation of the tensor modes propagation in this cosmological context.

In the framework of torsional gravity ones uses the tetrad fields
$e^\mu_A$, which form an orthonormal base in the tangent space of the underlying manifold 
$(\mathcal{M},g_{\mu\nu})$, where  $g_{\mu\nu}=\eta_{A B} e^A_\mu e^B_\nu$ is the 
metric tensor defined on
the manifold $\mathcal{M}$. Throughout this work we use the Greek indices to denote 
the coordinate space and the Latin indices for the tangent space. Furthermore, 
unlike general relativity which uses the torsionless Levi-Civita connection, here we 
  use the curvatureless
Weitzenb{\"{o}}ck connection $\overset{\mathbf{w}}{\Gamma}^\lambda_{\nu\mu}\equiv
e^\lambda_A\:\partial_\mu e^A_\nu$  \cite{Pereira.book}. Thus, 
the gravitational field is described by the torsion tensor
\begin{equation}\label{energy-momentum}
T^\lambda_{\verb| |\mu\nu} \equiv e^\rho_A
\left( \partial_\mu e^A_\nu - \partial_\nu e^A_\mu \right).
\end{equation} 
The Lagrangian of the teleparallel equivalent of general relativity (TEGR) is
the torsion scalar $T$,  constructed as
\cite{Pereira.book}
\begin{equation}
\label{T-scalar}
T\equiv\frac{1}{4}
T^{\rho \mu \nu}
T_{\rho \mu \nu}
+\frac{1}{2}T^{\rho \mu \nu }T_{\nu \mu\rho}
-T_{\rho \mu}{}^{\rho }T^{\nu\mu}{}_{\nu}\, \, ,
\end{equation}
and the corresponding  action reads
  $ S= \frac{1}{16 \pi G} \int d^4x e\, T $, where $e = \text{det}(e_{\mu}^A) = 
\sqrt{-g}$ and
$G$ is the Newton's gravitational constant (we set the speed of light to $c =1$). 

If we use TEGR as the starting point for torsional modified gravity, the simplest such 
modification is $f(T)$ gravity, with action 
\begin{align}
\label{action}
 S= \frac{1}{16 \pi G} \int d^4x e\, f(T) ~.
\end{align}
Variation of (\ref{action}) with
respect to the tetrads leads to the field equations 
\begin{align}
\label{field-eqs}
 & e^{-1}\partial_{\mu} (ee_A^{\rho}S_{\rho}{}^{\mu\nu}) f_{T} + e_A^{\rho}
S_{\rho}{}^{\mu\nu} \partial_{\mu}({T}) f_{TT} \nonumber\\
 & - f_{T} e_{A}^{\lambda} T^{\rho}{}_{\mu\lambda} S_{\rho}{}^{\nu\mu} + \frac{1}{4} e_
{A}^{\nu} f(
{T}) 
 = 4 \pi G e_{A}^{\rho} \Theta_{\rho}{}^{\nu} ~,
\end{align}
where $f_{T}=\partial f/\partial T$, $f_{TT}=\partial^{2} f/\partial T^{2}$, and 
$\Theta_{\rho}
{}^{\nu}$  denotes the energy-momentum tensor of the matter sector. In the above equation 
we have introduced for convenience the
``super-potential''  
$S_\rho^{\:\:\:\mu\nu} \equiv \frac{1}{2} \left( {\cal{K}}^{\mu\nu}_{\:\:\:\:\rho} +
\delta^\mu_\rho \: T^{\alpha\nu}_{\:\:\:\:\alpha} - \delta^\nu_\rho \:
T^{\alpha\mu}_{\:\:\:\:\alpha} \right) ~.$

Applying $f(T)$ gravity in a cosmological framework, namely  imposing the homogeneous 
and isotropic geometry
$e_{\mu}^A={\text
{diag}}(1,a,a,a)$,
which corresponds to the spatially
flat Friedmann-Robertson-Walker (FRW)  metric
\begin{equation}
ds^2= dt^2-a^2(t)\,  \delta_{ij} dx^i dx^j,
\end{equation}
with $a(t)$ the scale factor, and inserting it into the general field equations 
(\ref{field-eqs}), we extract the Friedmann equations 
\begin{eqnarray}\label{Fr11}
&&H^2= \frac{8\pi G}{3}\rho_m
-\frac{f}{6}+\frac{Tf_T}{3}\\\label{Fr22}
&&\dot{H}=-\frac{4\pi G(\rho_m+p_m)}{1+f_{T}+2Tf_{TT}}.
\end{eqnarray}
In the above equations $H\equiv\dot{a}/a$ is the Hubble function, with dots denoting
derivatives with respect to $t$, and $\rho_m$, $p_m$ are respectively the energy density 
and pressure for the matter perfect fluid.

Let us now study the   perturbations of $f(T)$ around an FRW cosmological background, 
focusing on the gravitational wave part. We follow \cite{Cai:2018rzd}
and we perturb the 
tetrads as 
\begin{equation}\label{decomposition}
 e^A_{\mu}(x) = \bar{e}^A_{\mu}(x) + \chi^A_{\mu}(x)~,
\end{equation}
where  $\bar{e}^A_{\mu}$ 
represents the part of the tetrad corresponding to metric components, 
which satisfies the equation $g_{\mu\nu}(x) = \eta_{AB} e^A_{\mu} e^B_{\nu} = \eta_{AB} 
\bar{e}^A_{\mu} \bar{e}^B_{\nu}$ (the part $\chi^A_{\mu}$
that represents the degrees of
freedom released from the local Lorentz transformation is not going to play any role in 
the analysis of this work). 
The  part $\bar{e}^A_{\mu}$ around a spatially flat FRW geometry 
writes as \cite{Chen:2010va,Wu:2012hs,Cai:2018rzd}
\begin{align}
\label{perturbation}
 \bar{e}^0_{\mu} =& \delta^0_{\mu} (1+\psi) + a \delta^i_{\mu} (G_i+\partial_i F) ~,
\nonumber \\
 \bar{e}^a_{\mu} =& a \Big[ \delta^a_{\mu}(1-\phi) \nonumber \\
  	& \ \ \ \, + \delta^i_{\mu} \delta^{aj} \Big( \frac{1}{2} h_{ij} + \partial_i
\partial_j B + \partial_j C_i + \partial_i C_j \Big) \Big] ~, \nonumber \\
 \bar{e}_0^{\mu} =& \delta_0^{\mu} (1-\psi) - \frac{1}{a} \delta^{\mu i} (G_i+\partial_i
F) ~, \nonumber \\
 \bar{e}_a^{\mu} =& \frac{1}{a} \Big[ \delta_a^{\mu} (1+\phi) \nonumber \\
  	& \ \ \ \, - \delta^{\mu i} \delta^j_a \Big( \frac{1}{2} h_{ij} + \partial_i
\partial_j B + \partial_i C_j + \partial_j C_i \Big) \Big] ~,
\end{align}
where $\phi$, $\psi$, $B$ and $F$ are the scalar modes, $C_{i}$ and 
$G_{i}$    the transverse vector modes, and 
$h_{ij}$ the transverse traceless tensor mode,
  which lead to the standard perturbed metric
\begin{align}
 g_{00} = & -1-2\psi ~, \nonumber\\
 g_{i0} = & -a [\partial_i F+G_i] ~,  \\
 g_{ij} = & a^2 [ (1-2\phi) \delta_{ij} + h_{ij} + \partial_{i} \partial_j B + \partial_j
C_i + \partial_i C_j ] ~.\nonumber
\end{align}
In the rest of the manuscript we set the   scalar and vector perturbations to   zero, 
since we are interested in studying the gravitational-wave sector.

Inserting (\ref{perturbation}) into (\ref{energy-momentum}) we acquire
\begin{align}
 T^i{}_{0j} = & H\delta_{ij} + \frac12 \dot{h}_{ij} \nonumber \\
 T^i{}_{jk} = & \frac12 \left( \partial_j h_{ik} - \partial_k h_{ij} \right) ~,
\end{align}
and consequently the torsion scalar (\ref{T-scalar}) is perturbed as
\begin{align}
T =T^{(0)}+O(h^2)= 6H^2 +O(h^2),
\end{align}
with $T^{(0)}$ the zeroth-order quantity. Thus, we deduce that at  linear order the 
torsion scalar remains unaffected, which lies behind the fact that in $f(T)$ gravity the 
gravitational waves do not have extra polarization modes \cite{Farrugia:2018gyz}. 
Moreover,
the perturbed super-potential  can be written as 
$ S_i{}^{0j} = H\delta_{ij} -\frac14\dot{h}_{ij}$ and $
 S_i{}^{jk} = \frac{1}{4a^2} (\partial_jh_{ik} -\partial_k h_{ij})$.
Inserting the perturbed quantities into the field equations  
(\ref{field-eqs}), and neglecting the matter sector, we obtain \cite{Cai:2018rzd}
\begin{align}
 & 4 f_T \Big[ \big( \dot{H}+3H^2 \big) \delta_{ij}+\frac14 \big(
-\ddot{h}_{ij}+\frac{\nabla^2}{
a^2}{h}_{ij}-3H\dot{h}_{ij} \big) \Big] \nonumber \\
 & + 4\dot{f}_T \big( H\delta_{ij}-\frac{\dot{h}_{ij}}{4} \big) -f\delta_{ij}=0 ~,
\end{align}
where the derivative $f_{T}$ is calculated at
$T=T^{(0)}$. Hence,  the perturbation part of the above 
equation leads to the
equation of motion for the gravitational waves in $f(T)$ cosmology, namely
\begin{equation}
\label{gw_ft}
h''_{ij} + 2  \mathcal{H}(1 - \beta_T) h'_{ij} + k^2 h_{ij} = 0,
\end{equation}
with primes denoting derivative with respect to the conformal time, and where  we 
have introduced the dimensionless parameter  \cite{Cai:2018rzd}
\begin{equation}
\label{beta_T}
\beta_T = - \frac{\dot{f}_T}{3 H f_T}.
\end{equation}
Therefore, we straightforwardly deduce that 
the  speed of GWs is equal to one, i.e. equal to the speed of light, and thus the 
experimental constraint of GW170817 is trivially satisfied in $f(T)$ 
gravity. However,  as it was mentioned in \cite{Cai:2018rzd}, the correction term 
$\beta_T$ reflects the effect on the gravitational waves due to the change that $f(T)$ 
gravity brings on the background they propagate on. 

In order to quantify the above effect we follow the analysis of 
\cite{Atsushi,Ezquiaga:2017ekz}. In most of gravitational modifications one can bring the 
GW propagation equation in the form 
\begin{equation}
\label{gw_general}
h''_{ij} + (2+\nu)  \mathcal{H}  h'_{ij} + (c_T^2  k^2+a^2 \mu^2) h_{ij} = \Pi_{ij},
\end{equation}
where  $c_T$ is the GW propagation 
speed, $\mu$ is the effective graviton mass, $\nu$ is related to the effectively 
running Planck mass, and $ \Pi_{ij}$ is a source term arising from possible additional 
fields. Hence, one can describe the 
deviation of GW propagation at cosmological scales comparing to general relativity 
through  
\begin{equation}
\label{basicmodif}
h = e^{-\mathcal{D}} e^{-ik \Delta T} h_{GR},
\end{equation}
where  
\begin{equation}
\label{damping_gw}
\mathcal{D} = \frac{1}{2} \int \nu \mathcal{H} d \tau'
\end{equation}
quantifies the amplitude modification, i.e. the damping factor, while 
\begin{equation}
\label{fase_gw}
\Delta T = \int \Big( 1 - c_T - \frac{a^2 \mu^2}{2 k^2} \Big) d\tau'
\end{equation}
quantifies the phase modification, i.e. the time delay. 

In the case of $f(T)$ gravity Eq. (\ref{gw_ft}) implies that $c_T = 1$, $\mu=0$,   $ 
\Pi_{ij}=0$ and $\nu=-2\beta_T$. Therefore, we deduce that in $f(T)$ gravity $\Delta T = 
0$, and thus we do not obtain any phase modification in comparison to general relativity. 
However, the quantity $\nu$, which carries the information of the $f(T)$ modification, 
will lead to a non-zero $\mathcal{D}$, and thus to an amplitude modification of the 
gravitational waves comparing to general relativity. Hence, this modification can in 
principle be measured in GW observations, offering an observational signature of this 
class of modified gravity. In the following   sections we quantify this behavior.

\section{Primordial gravitational waves}
\label{PrimordialGW}

In this section we investigate the deviation of GW propagation in $f(T)$ gravity, 
comparing to general relativity, focusing on the primordial gravitational waves. As a 
specific model, and without loss of generality, we will consider the power-law one, which 
is the most viable one, nevertheless our analysis can be performed for every $f(T)$ form. 
The power-law scenario corresponds to  \cite{Bengochea:2008gz}:
\begin{equation}
\label{the_model}
f(T) = T + \alpha(-T)^b.
\end{equation}
with $\alpha$ and $b$ the two parameters.
Inserting this  
into  (\ref{Fr11}) at present time we
obtain
\begin{eqnarray}
\alpha=(6H_0^2)^{1-b}\left(\frac{1-\Omega_{m0}-\Omega_{r0}}{2b-1}\right),
\end{eqnarray}
with $\Omega_{m0}$, $\Omega_{r0}$ respectively the current values of the matter and 
radiation density parameters, and $H_0$ the Hubble parameter at present, and thus the 
only free model parameter is 
$b$. The value $b=0$ corresponds to recovery of general relativity and of $\Lambda$CDM 
cosmology (in this case the parameter $\alpha$ is related to the cosmological constant).

We insert the above power-law $f(T)$ form into the modified propagation equations of 
the previous section, and we use the publicly available CLASS (Cosmic Linear 
Anisotropy Solving System) code \cite{class} in order to investigate the effects of the 
gravitational waves on the CMB BB anisotropy and also to calculate the propagation of the 
gravitational waves through cosmic time.

\begin{figure}[t]
\centering
\includegraphics[width=3.25in, height=2.5in]{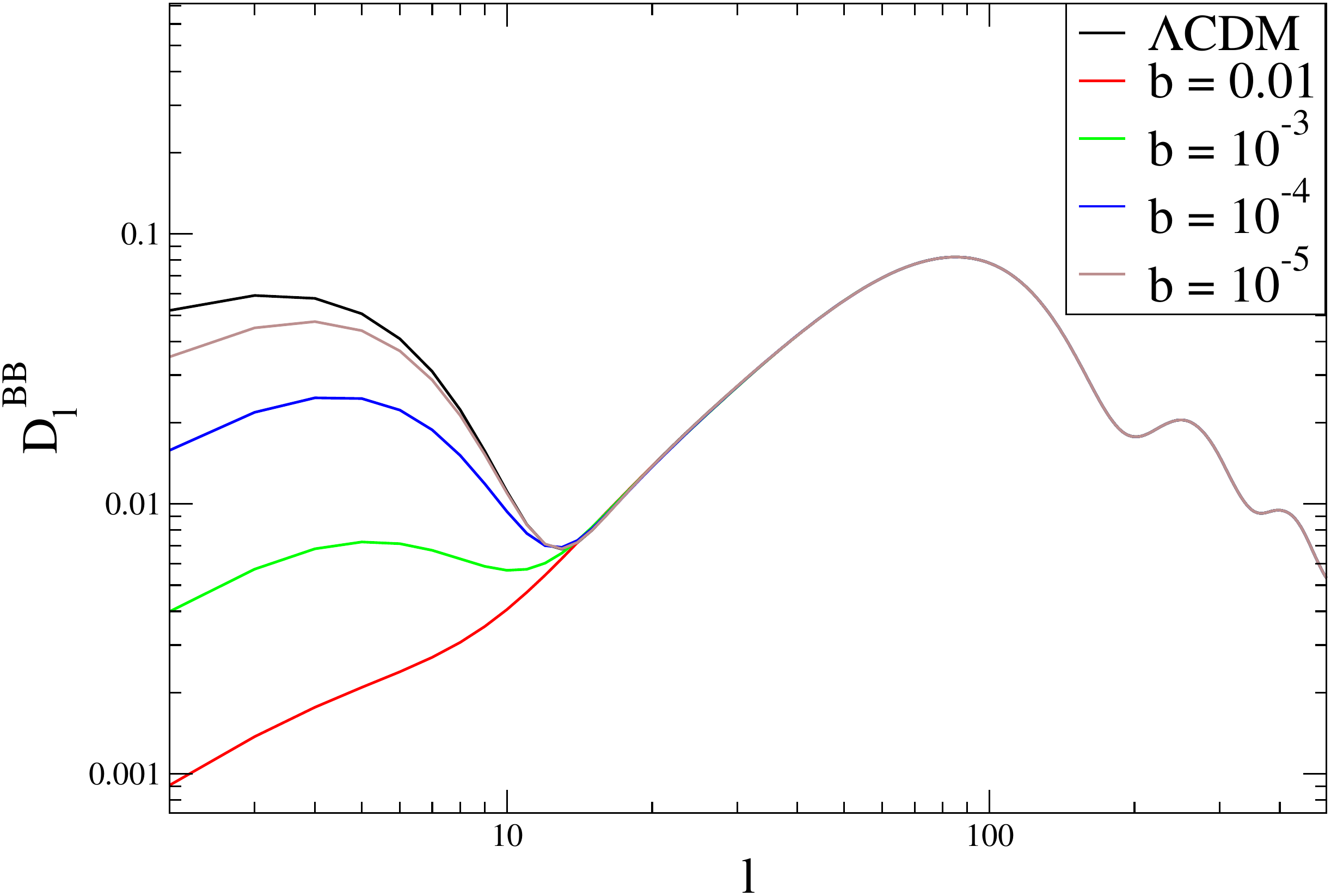}
\caption{{\it{The  CMB BB power spectrum, $D_l^{BB}= l(l+1)C_l^{BB}/2 
\pi \mu K $, for the $\Lambda$CDM cosmology and the $f(T)$ gravity power-law model 
(\ref{the_model}), for various values of the free model  parameter $b$.}}
}
\label{cmb_BB}
\end{figure}
In Fig. \ref{cmb_BB} we show the CMB BB spectrum for the $f(T)$ gravity and the 
$\Lambda$CDM cosmological model as the reference model, for tensor modes only. 
In drawing the graphs we have considered various values of the free model  parameter $b$. 
All other parameters are fixed in the same way for both models, based on Planck 
2015 results \cite{Ade:2015xua}. As we observe, for the angular scale $l > 20$, the 
theoretical predictions for $f(T)$ gravity and $\Lambda$CDM are practically 
identical. Thus, for small angular scales no deviations are expected compared to   
standard $\Lambda$CDM cosmology. On the other hand, we can see significant deviations at 
large angular scales.  

Additionally, the theoretical CMB BB spectrum should also 
present a peak at $l \simeq 5$, still to be detected by   future experiments, due to 
the effects of tensor modes on the scattering during the reionization epoch. At low-$l$, 
the large angular is dominated by modes that have not entered the horizon at 
recombination 
and therefore are approximately constant. Thus, the effects due to $f(T)$ modification 
can be quantified on 
the reionization peak (which should be located at $l \simeq 5$), where we can note 
different predictions for a range of values of model parameter $b$ compared to the 
reference $\Lambda$CDM scenario. Interestingly, future measurements of the reionization 
peak by CMB spectrum could be used to discriminate between $f(T)$ gravity and standard 
cosmology scenario.

\begin{figure}
\includegraphics[width=3.0in,height=2.5in]{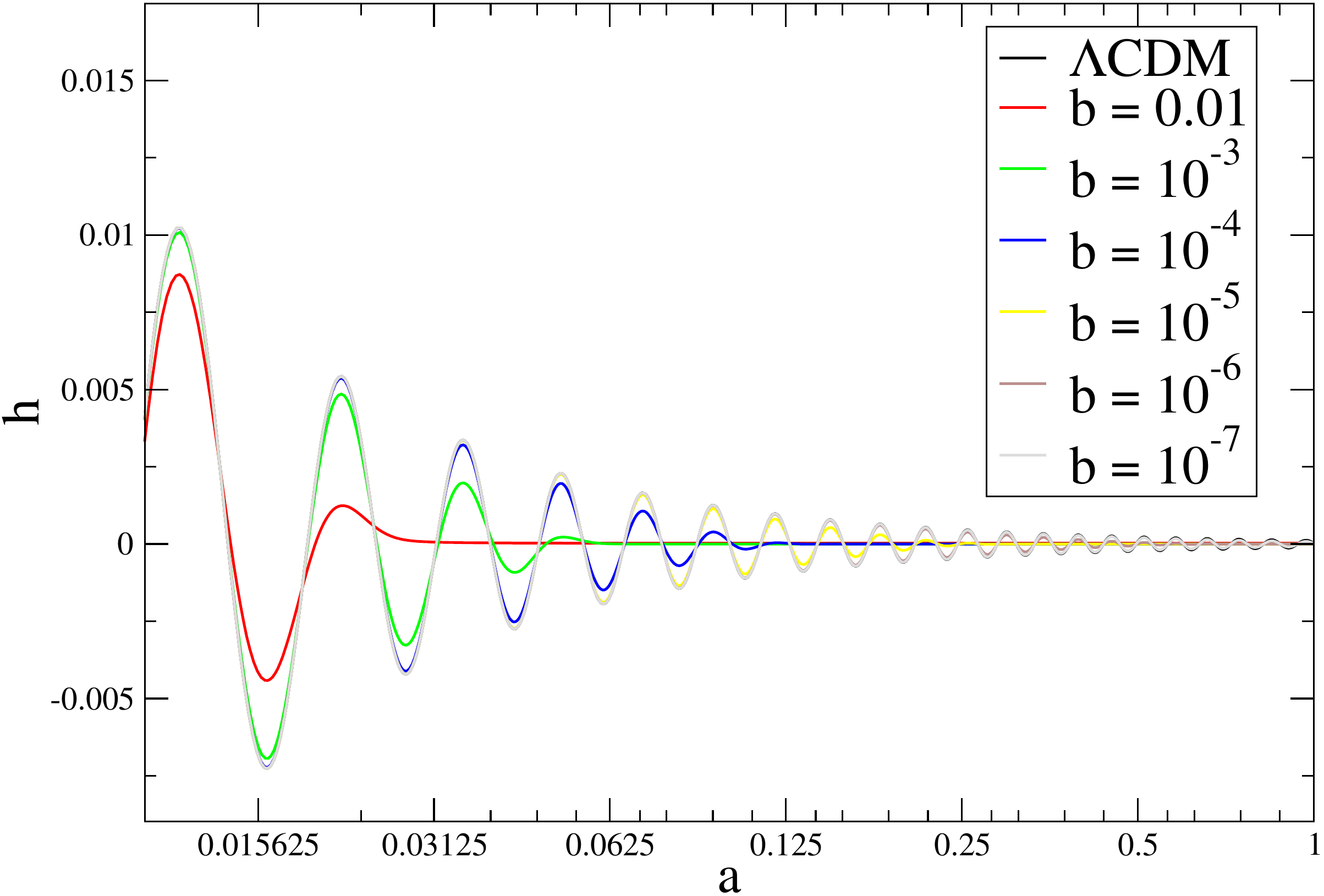}\quad \quad 
\includegraphics[width=3.0in,height=2.5in]{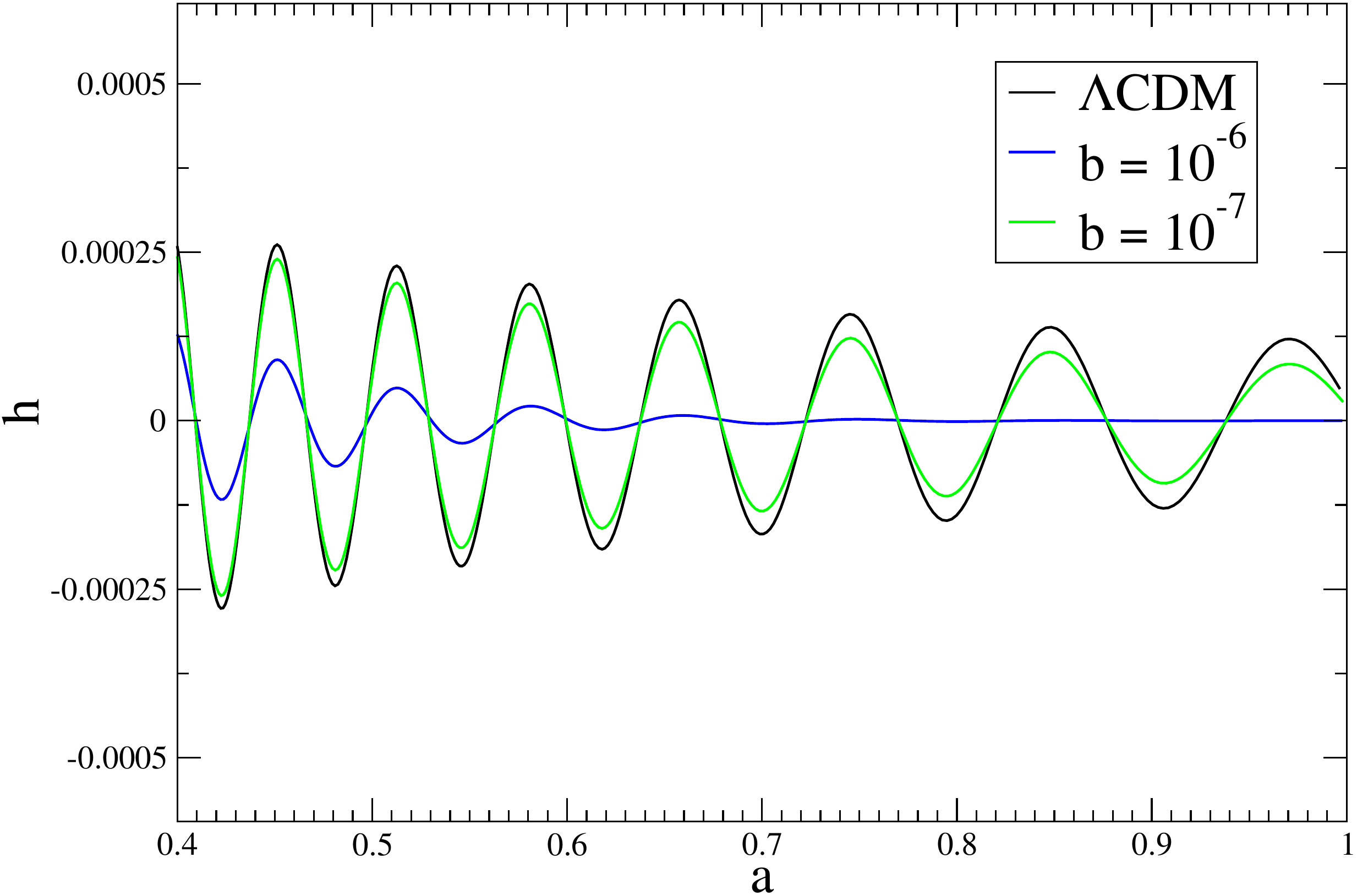}
\caption{{\it{Propagation of the gravitational waves at $k = 
0.01 {\rm Mpc^{-1}}$, in $\Lambda$CDM 
scenario and in the $f(T)$ gravity power-law model 
(\ref{the_model}) for various values of the free model  parameter $b$,  as a 
function of the scale factor $a = (1+z)^{-1}$. Upper graph: full cosmological 
history. Lower graph: late-time   cosmological 
history.}}}
\label{h_fT}
\end{figure}

In Fig. \ref{h_fT}, following the approach of \cite{Atsushi}, we depict the propagation 
of GWs as the function of the scale factor, for various values of the model parameter $b$ 
of the power-law form (\ref{the_model}), in order to investigate how $b$ affects the 
amplitude damping. It is already known that the amplitude of GWs decays rapidly 
immediately after the tensor-modes entry into the horizon, while before the 
entrance to the horizon the amplitude is practically constant. The expansion of the 
Universe leads to damping of the tensor modes with the term proportional to $h'$. We have 
noted that for $a < 0.01$, namely at early times, taking different ranges in orders of 
magnitude of $b$, only a tiny deviation from  $\Lambda$CDM paradigm appears.
On the other hand, for $a > 0.01$ the damping factor becomes significant, and we can see 
that already at the matter era the amplitude of GWs becomes practically null (or 
extremely small) for  $b>10^{-5}$. For instance, for $b= 10^{-2}$
 we have $h \simeq 0$ at $a 
\simeq 0.04$, while for $b=10^{-5}$ we get that $h \simeq 0$ at $a 
\simeq  0.4$. Moreover, when the Universe enters into the accelerated phase 
only for the cases $b= 10^{-6}, \, 10^{-7}$  the tensor-modes oscillations  are non-null, 
however   in the case $b= 10^{-6}$ the modes quickly decay and only in the case $b= 
10^{-7}$ they survive up to present time.
Concerning the phase, we do not observe any difference, since as we discussed there is no 
phase modification in $f(T)$ gravity ($\Delta T = 0$ in (\ref{basicmodif})). 

In summary, the more $b$ departs from its general relativity  value $b=0$ the larger is 
the GW amplitude decay comparing to $\Lambda$CDM scenario. This is because for larger $b$ 
values the tensorial modes enter in the cosmological horizon earlier, compared to the 
standard prediction within $\Lambda$CDM cosmology, and thus the GW amplitude goes 
rapidly to zero already in radiation and matter eras. 
Hence, we deduce that a possible future detection of primordial gravitational waves 
would imply that $ b \lesssim 10^{-7}$ (taking the specific scale $k = 0.01 {\rm 
Mpc^{-1}}$), 
bringing the viable $f(T)$ gravity models five 
orders of magnitude closer to $\Lambda$CDM cosmology comparing to standard cosmological 
constraints based on SN Ia, BAO, CMB, $H(z)$   data
\cite{Wu:2010mn,Nesseris:2013jea,Nunes:2016qyp,Nunes:2016plz,Nunes:2018xbm,
Basilakos:2018arq}. This 
fact reveals the 
capabilities of gravitational-wave astronomy, since such strong constraints were possible 
to be obtained only through Solar-System data \cite{Iorio:2012cm}.

\section{Planck and CORE constraints}
\label{PlanckCORE}

In this section we present new observational constraints on $f(T)$ gravity, arising 
from the CMB data from Planck collaboration \cite{Ade:2015xua}, as well as from forecasts 
that are expected from near-future probes such as CORE 
\cite{DiValentino:2016foa,Finelli:2016cyd}. Similarly to the previous section we will 
use the CLASS code \cite{class}. We mention that in the present analysis we only 
consider the CMB data, neglecting the influences of the external data, 
since we are particularly interested in quantifying and analyzing the effects of 
equations 
(\ref{gw_ft})-(\ref{beta_T}), i.e. the modified GWs propagation in $f(T)$ gravity, as 
well as its effects on CMB. The $f(T)$ 
gravity has been recently well constrained from the geometrical data 
\cite{Nunes:2016qyp,Basilakos:2018arq}, and using 
the CMB data for the first time in \cite{Nunes:2018xbm}.
 
Concerning the Planck data we use the likelihood {\it fake planck realistic} included 
in MontePython code \cite{montepython}, taking into account the temperature, polarization 
and CMB lensing extraction, and we adopt noise spectra roughly matching those 
expected from the full Planck results. The baseline parameter space is given by
\[ \mathcal{P} \equiv \{100 \omega_{\rm b}, \, \omega_{\rm cdm}, \, \ln10^{10}A_{s}, \,  
 n_s, \, \tau_{\rm reio}, \, H_0, \, r, \, b \},\]
where the parameters in $\mathcal{P}$ from left to right 
are respectively the  baryon density, the cold dark matter 
density, the amplitude and slope of the primordial spectrum of metric fluctuations, the
scalar spectral index, the optical depth to reionization, the Hubble constant, the 
tensor-to-scalar 
ratio, and the free parameter (i.e. $b$) of the power-law $f(T)$ model 
(\ref{the_model}). The 
priors used for the model parameters are   the following:
$\omega_{\rm b} \in [0.005, 0.1]$, $\omega_{\rm cdm} \in [0.01, 0.99]$, 
$\ln10^{10}A_{s} \in [2.4, 4.0]$, $n_s \in [0.5, 1.5]$, 
$\tau_{\rm reio} \in [0.01, 0.8]$, $H_0 \in [50, 90]$, $r \in [0, 1.0]$, and
$b \in [0, 0.1]$.
 
Concerning the CORE analysis we use the 
likelihood {\it CORE m5} also included in MontePython. The experimental 
specifications for CORE data in our analysis are summarized in Table \ref{core_data}.

\begin{table}[ht]
\begin{center}
\begin{tabular}{c|c|c|c}
\hline
Channel [GHz]  &  FWMH [arcmin] &  $\Delta T$ & $\Delta P$  \\
\hline
{130} & 8.51 & 3.9 & 5.5 \\

{145} & 7.68 & 3.6 & 5.1 \\

{160}  & 7.01 & 3.7 & 5.2  \\

{175}  & 6.45 & 3.6 & 5.1 \\

{195} & 5.84 & 3.5 & 4.9 \\

{220} & 5.23 & 3.8 & 5.4  \\
\hline
\end{tabular}%
\end{center}
\caption{\label{core_data} Experimental specifications for CORE, with frequency channels 
dedicated to cosmology, and beam width, temperature sensitivity, and polarization 
sensitivity for each channel, in units of $\mu$K arcmin.}
\end{table}

In our forecasts, we assume $l_{\rm min} = 2$, $l_{\rm max} = 3000$, and $f_{\rm sky} = 
0.70$. We forecast the $f(T)$ gravity with the set of cosmological parameters shown in 
$\mathcal{P}$.

In the forecasting analysis we assume the fiducial values of the above parameters as: 
\{2.22, 0.119, 3.07, 0.962, 0.05, 68.0, 0.1, 0.005\}. The details of the methodology used 
in the CORE likelihood can be seen in \cite{DiValentino:2016foa,Finelli:2016cyd}.

\begin{figure}[ht]
\includegraphics[width=3.25in, height=3.25in]{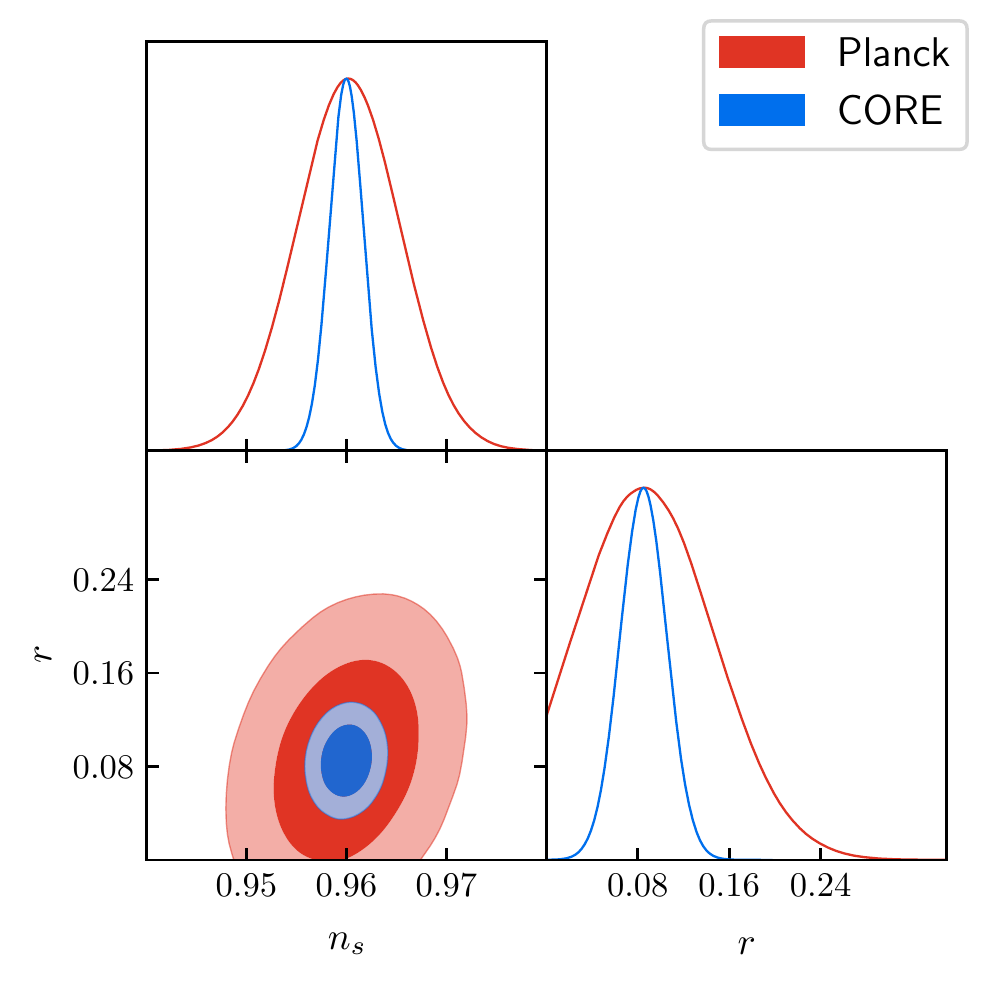} 
\includegraphics[width=3.25in, height=3.25in]{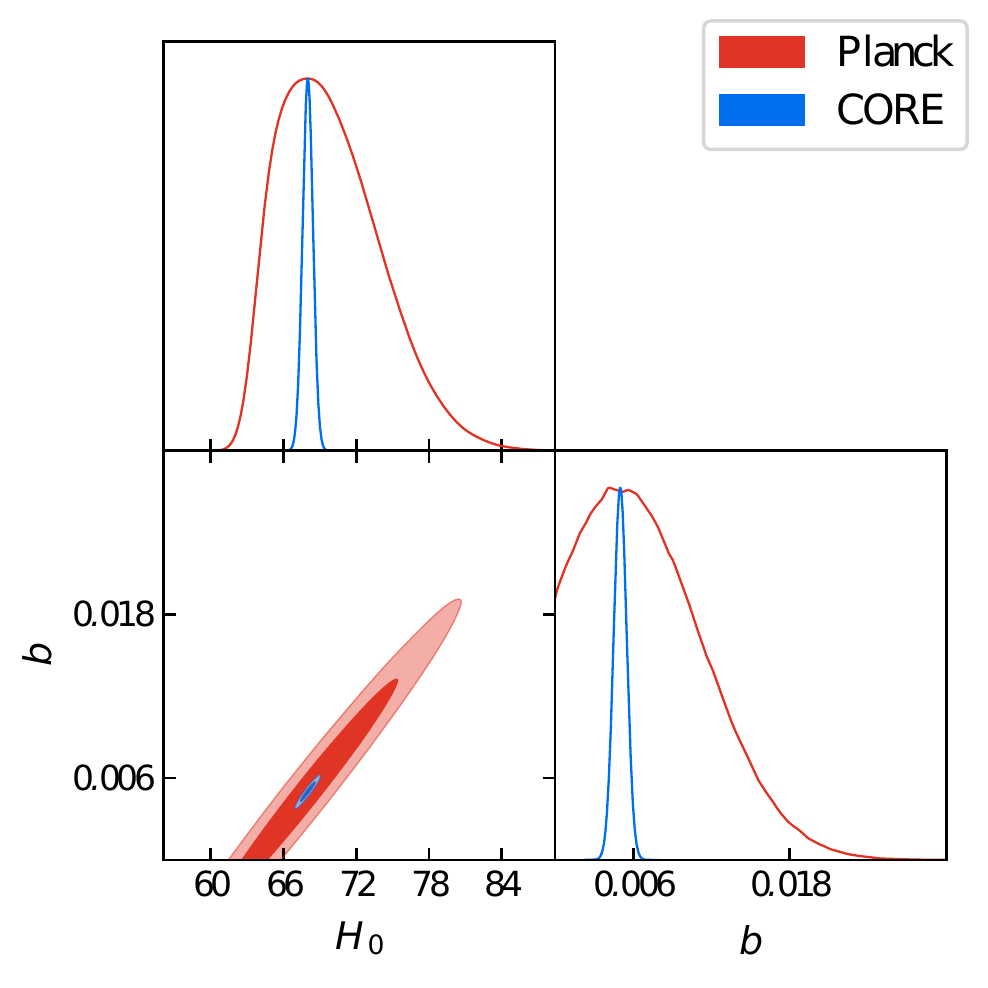} 
\caption{ 
\label{rns} 
{\it{ Upper graph: One-dimensional marginalized posterior 
distributions and 68\%, 95\% CL regions in the $n_s-r$ plane at $k = 0.05 \, {\rm 
Mpc^{-1}}$, from the Planck data and CORE 
forecasts.  Lower graph: One-dimensional marginalized posterior distributions and 
68\%, 95\% CL regions in the $H_0-n$ plane (where $b$ is the free parameter of the $f(T)$ 
gravity 
model (\ref{the_model})) from the Planck data and CORE 
forecasts.}}}
\end{figure}
 
In the upper graph of Fig. \ref{rns} we present the parametric space in the 
$n_s - r$ plane, where for the analysis we have assumed the pivot scale at $k = 0.05 \, 
{\rm Mpc^{-1}}$, which is the standard in 
the literature. As we observe, $r < 0.23$ at 95\% confidence level (CL) and $n_s = 0.96 
\pm 0.005$ at 68\% CL from the Planck data. These results are also in accordance with 
the Planck team  assuming $\Lambda$CDM \cite{Planck_inflation_2018}, where no evidence 
for a non-null $r$ is also reported and $n_s < 1$ up to $5 \sigma$ CL. According to the 
authors' knowledge, this is the first observational constraints on inflation parameters 
(plane $n_s - r$) obtained directly in the context of $f(T)$ gravity.

In the lower graph of Fig. \ref{rns} we present the parametric space in 
the $H_0 - b$ plane. As we can see, $H_0 = 69 \pm 5.42 $ km/s/Mpc and $b = 0.005 \pm 
0.006 $, both at 68\% CL. As already noted in \cite{Nunes:2018xbm} (within a different 
theoretical and observational perspective), $f(T)$ gravity can provide high $H_0$ values, 
and therefore a possible solution to the $H_0$ tension might be available in this context 
\cite{Nunes:2018xbm}.

In this work, incorporating the propagation modification of the tensor modes due to 
$f(T)$ gravity and its effects on CMB, we can again see that 
$H_0$ is fully compatible with local measurements, namely
$H_0 = 73.2 \pm 1.74$ \cite{Riess}, even at 1$\sigma$ CL. 

We proceed motivated by the future CMB experiments, and we apply the inherent 
modifications due to $f(T)$ gravity aiming to impose   bounds on the baseline 
parameters 
of the model. In the upper panel of Fig. \ref{rns} we additionally present the 
parametric space in the plane $n_s - r$ via CMB CORE forecasts. In this way we can see 
how much can CORE improve the constraints compared to the current Planck ones.
We find $r = 0.1 \pm 0.020 $ and $n_s = 0.96 \pm 0.0017 $ at 68\% CL (as we mentioned 
above, we assume $r = 0.1$ and $n_s = 0.96$ in the forecast data simulation). 
Let us define the improvement with respect to the constraints
arising from the Planck data as $I_{i} = \sigma_{Planck}/\sigma_{CORE}$ 
\footnote{The label 
$i$ runs over the parameters given in $\mathcal{P}$.}. 
Thus, for the parameters of our interest, we find that $I_{r} = 11.5$ and $I_{n_s} = 
2.94$.
Within the $\Lambda$CDM paradigm, the CORE collaboration 
\cite{Finelli:2016cyd} finds for instance $r = 0.0042 \pm 0.00028$, assuming $r = 0.0042$  
in the 
forecast analysis. For other forecastings using CORE estimation see 
\cite{DiValentino:2016foa,Finelli:2016cyd}. 

In the lower graph  of Fig. \ref{rns} we depict the bounds that CORE results can impose 
directly on the free model parameter of the power-law $f(T)$ model, and in particular we 
find that $b = 0.0050 \pm 0.00049$. Here, we note an improvement of $I_{b} = 
12.2$. As we 
can see, future constraints could 
improve current constraints with respect to the Planck data, on the $f(T)$ gravity free 
parameter, by a 
factor of $12$. A precise measurement of this parameter, 
with that magnitude of improvement,
can offer an opportunity to check  deviations from the general theory of relativity.
The value $b = 0.0050$ is the best-fit value that arises from the Planck data too, thus 
it 
is reasonable to consider it for performing our forecast simulation. Hence, we may 
conclude that the forecast errors on $b$ 
may limit a non-null value on this parameter by taking future CMB missions. This will be 
a clear signature that a deviation from $\Lambda$CDM cosmology is favored.

\section{Gravitational waves from mergers}
\label{mergersGWs}

In the previous section we investigated the effects of $f(T)$ gravity on the primordial 
GWs. In this section we present a preliminary discussion on the  possibility to use the
properties of the detected  GWs that arise from mergers, in order to impose constraints 
on 
$f(T)$ gravity and on modified gravity in general. In this direction, the GWs that are 
accompanied by electromagnetic counterparts are proved to be extremely efficient.

In the case of detection of GWs that arise from black holes mergers ones does not have 
any other information apart from the properties of the GWs at the moment they reach 
earth, and the direction in space they came from (in the case where three different 
detectors are used). Hence, assuming that general relativity is the underlying 
gravitational theory and that $\Lambda$CDM cosmology is the background cosmological model 
one can calculate the distance of the merger, the involved black hole masses, and the 
properties of the GWs at the moment of emission 
\cite{ligo01,Abbott:2016nmj,Gw03,Gw04,Gw05,Gw06}. 
Definitely, assuming a modified gravity and its implied cosmology as the underlying 
theory will lead to different calculations, and thus one faces degeneracies that do not 
allow for distinguishing different theories.

However, the situation changes radically in the case of detection of GWs that arise from 
neutron stars mergers, which are accompanied by an electromagnetic counterpart. In this 
case apart from the properties of the GWs at the moment they reach earth 
and their origin direction, one can additionally calculate their propagation 
speed, the distance of the merger, and the involved neutron stars masses, independently 
of the underlying gravitational theory and cosmological scenario, just using the implied 
physics from the electromagnetic information \cite{Gw07,GRB17}. Therefore, without 
assuming an underlying cosmology one can find the properties of the GWs at the time of 
their emission too. Hence, knowing the GWs properties at both the emission and detection 
time one can extract information for the background on which they propagated and thus 
impose constraints on the gravitational theory that determines it. 
In the case of $f(T)$ gravity, as we described in Section \ref{sec-model}, the effect of 
the $f(T)$ modification on the underlying cosmology leads to an amplitude modification 
comparing to the GWs propagating in $\Lambda$CDM scenario under general relativity, 
without a phase change. 

In principle one can follow the above roadmap and impose new constraints on $f(T)$ 
gravity. In practice however, the involved procedure is very complicated and one needs to 
perform a thorough investigation on the GWs generation from neutron star mergers 
in the framework of $f(T)$ gravity itself. In these lines the analysis can be based on 
the recent work \cite{GW_MG05}, focusing on the strain amplitude of the generated Gws. 
Since the necessary calculations are lengthy and complicated and lie beyond the scope of 
the present work, which is  mainly the investigation of cosmological GWs, it is left for 
a separate project.

\section{Conclusions}
\label{Conclusions}

The advancing GWs multi-messenger astronomy offers a new window to observe nature, and 
amongst others to extract information about cosmology and gravity. In this work we 
focused on the case of $f(T)$ gravity, and we studied the new observational constraints 
that arise from  the effect of primordial GWs  on the cosmic microwave background (CMB) 
anisotropies and the BB spectrum.

The main feature of the investigation that lies at the basis of the analysis, is that the 
underlying gravitational theory determines both the properties of the GWs themselves 
(speed, polarization modes, etc) as well as the properties of the background on which 
they propagate (the expanding universe). Hence, one can use the GW features in order to 
impose constraints on the various gravitational  theories and offer a way to distinguish 
them. In the particular case of $f(T)$ gravity (where it was recently shown that GWs 
propagate with the light speed \cite{Cai:2018rzd,Li:2018ixg}, without extra polarization 
modes \cite{Bamba:2013jqa}), we showed that one obtains only an amplitude modification 
and 
not a phase one on the GWs propagation, comparing to the case of general relativity in 
the 
background of $\Lambda$CDM cosmology.

Concerning primordial GWs, and focusing without loss of generality on the power-law 
$f(T)$ model, we showed that  the more the model departs from general relativity, i.e. 
the more the exponent $b$ departs from its general relativity value $b=0$, the larger is 
the GW amplitude decay comparing to $\Lambda$CDM scenario. Hence, we deduced that a 
possible future detection of primordial gravitational waves 
would imply that $ b \lesssim 10^{-7}$ (analyzing on the scale $k = 0.01 \, {\rm 
Mpc^{-1}}$), bringing the viable $f(T)$ gravity models five 
orders of magnitude closer to $\Lambda$CDM cosmology comparing to standard cosmological 
constraints based on SN Ia, BAO, CMB, $H(z)$  data
\cite{Wu:2010mn,Nesseris:2013jea,Nunes:2016qyp,Nunes:2016plz,Nunes:2018xbm,
Basilakos:2018arq}. This fact reveals the capabilities of gravitational-wave astronomy, 
since such strong constraints were possible to be obtained only through Solar-System data 
\cite{Iorio:2012cm}.

Additionally, we used the CLASS code in order to quantify the primordial GWs effect on 
the CMB anisotropies and the BB spectrum. We used both the data from the Planck 
probe, as well as forecasts from the near-future CORE collaboration. As we showed,   
possible non-trivial constraints on the tensor-to-scalar ratio would favor a model 
parameter $b$ different from its general relativity value, offering a clear signature of 
$f(T)$ gravity comparing to $\Lambda$CDM cosmology.

Finally, we examined the  possibility of constraining $f(T)$ gravity through the 
detection of GWs that arise from neutron stars mergers, which are accompanied by an 
electromagnetic counterpart. In this case, apart from observing the properties of the GWs 
at the time of detection one can use the features of the electromagnetic observations and 
the deduced physics of the neutron stars to find the properties of the GWs at the time of 
their emission, without any assumption on the underlying theory of gravity and the 
background cosmological evolution. Hence, knowing the GWs properties at both the emission 
and detection time it is possible to extract information for the background on 
which they propagated and thus impose constraints on the gravitational theory that 
determines it. Since the basic scope of the present work is mainly the investigation of 
cosmological Gws, the above detailed investigation is left for a separate project.

In summary, we showed how one can use information from the advancing multi-messenger GW 
astronomy in order to extract new observational constraints on $f(T)$ gravity. As we 
saw, $f(T)$ gravity remains in agreement with observations and thus a good candidate 
for the description of nature.

\begin{acknowledgments}
\noindent The authors thank the referee for some specific clarifying points.  
The authors would also like to thank  J. C. N. de Araujo and  M. E. S. Alves for
useful discussions. This article is based upon 
work from CANTATA COST (European Cooperation in Science and Technology) action CA15117, 
EU 
Framework Programme Horizon 2020.
\end{acknowledgments}

\end{document}